
\magnification=\magstep1
\def\sp{\,\,\,} \overfullrule=0pt
 \def\h{h^\vee} \def\c{\chi} \def\C{\chi^*}  
\def\T{{\cal T}}   \def\Q{{\bf Q}}
\def\l{\Lambda}   \def\la{\lambda}   \def\R{{\bf R}} \def\Z{{\bf Z}}
\def\u{\tau}       \def\equi{\,{\buildrel \rm def \over =}\,}
\def\eg{{\it e.g.}$\sp$} \def\ie{{\it i.e.}$\sp$}
 \def\g{{\hat g}}

{\nopagenumbers
\rightline{August, 1992}
\bigskip \bigskip
\centerline{{\bf Partition Functions for Heterotic}}\bigskip
\centerline{{\bf WZW Conformal Field Theories}}
\bigskip \bigskip
\centerline{Terry Gannon}
\centerline{{\it Mathematics Department, Carleton University}}
\centerline{{\it Ottawa, Ont., Canada K1S 5B6}}\bigskip \bigskip \bigskip
Thus far in the search for, and classification of, `physical' modular
invariant partition functions $\sum N_{LR}\,\c_L\,\C_R$ the attention
has been focused on the {\it symmetric} case where the holomorphic and
anti-holomorphic sectors, and hence the characters
$\c_L$ and $\c_R$, are associated
 with the same Kac-Moody algebras $\g_L=\g_R$ and levels $k_L=k_R$.
In this paper we consider the more general possibility where $(\g_L,k_L)$ may
not equal $(\g_R,k_R)$. We discuss which choices of algebras and levels may
correspond to well-defined conformal field theories, we
find the `smallest' such  {\it heterotic} (\ie asymmetric) partition
functions, and we give a method, generalizing the Roberts-Terao-Warner
lattice method,
for explicitly constructing many other modular invariants.
We conclude the paper by proving that this new lattice method will succeed in
 generating all the heterotic partition functions, for all choices of algebras
and levels.
\vfill \eject} \pageno=1

 \centerline{{\bf 1. Introduction}} \bigskip

Rational conformal field theories [1] consist of two nearly independent
sectors,
the holomorphic (``left-moving'') and anti-holomorphic (``right-moving''),
coupled via the partition function. The currents $J_L(z)$, $J_R(z^*)$
corresponding to the two sectors of a Wess-Zumino-Witten theory [2] generate
two commuting Kac-Moody algebras ${\hat g}_L,{\hat g}_R$. Its partition
function can be written as:
$$Z(z_Lz_R|\u)=\sum N_{{\hat \lambda}_L {\hat \lambda}_R} \c_{{\hat \lambda}_L}
(z_L,\u)\,\c_{{\hat \lambda}_R}(z_R,\u)^*, \eqno(1.1)$$
where $\c_{{\hat \lambda}_L}$ is the character of the representation of ${\hat
g}_L$ with highest weight ${\hat \lambda}_L$ (similarly for
$\c_{{\hat \lambda}_R}$), the coefficients
$N_{{\hat \lambda}_L {\hat \lambda}_R}$ are numbers (multiplicities), and
the sum is over all highest weights with the levels $k_L,k_R$ fixed by
the theory. All this will be described more carefully in the next section.

The algebras $\g_L$, $\g_R$ may or may not be isomorphic, and $k_L$, $k_R$
may or may not be equal.
There are neither physical nor mathematical reasons why the
case $(\g_L,k_L)\neq (\g_R,k_R)$ should be avoided,
 and in fact experience from string theory [3] hints that this case could
include some interesting theories.
Any such asymmetric theory, and its partition function, shall be called
{\it heterotic}.

There are three properties the sum in eq.(1.1) must satisfy in order
to be the partition function of a physical conformal field theory:
\item{(P1)} it must be {\it modular invariant}. This is equivalent to
the two conditions:
$$\eqalignno{&Z(z_Lz_R|\u+1)=Z(z_Lz_R|\u), &(1.2a)\cr
\exp[-\pi i(k_Lz_L^2/\u-k_Rz_R^{*2}/\u^*)]\, &
Z(z_L/\u, z_R/\u|-1/\u)=Z(z_L z_R|\u); &(1.2b)\cr}$$

\item{(P2)} the coefficients $N_{ \lambda_L \lambda_R}$
in eq.(1.1) must be {\it non-negative integers}; and

\item{(P3)} {\it uniqueness of the vacuum}: $N_{{\hat \la}_L{\hat \la}_R}=1$
for ${\hat \la}_L=k_L{\hat \beta}_{0L}$, ${\hat \la}_R=k_R{\hat \beta}_{0R}$
(these are the highest weights of the singlet
 representations of levels $k_L$, $k_R$, respectively; ${\hat \beta}_{0L}$ and
${\hat \beta}_{0R}$ are the fundamental weights (see [10]) associated with
the 0-nodes of the Dynkin diagrams of $\g_L$ and $\g_R$, respectively).
In the following sections we switch
notation from affine weights ${\hat \la}$ to horizontal weights $\la$, and
this condition reduces to $N_{00}=1$.

If the function $Z$ in (1.1) satisfies (P1), we will call it an {\it
invariant};
 if in addition it satisfies (P2) and (P3) we will call it a {\it physical
invariant}. These properties are necessary for $Z$ to be the partition
function of a sensible theory, but they are not sufficient. In this paper
we will be interested in the construction and classification of all
physical invariants corresponding to a given choice of algebras and levels,
but we will not address the question of which of these actually correspond
to well-defined conformal field theories.

Much work has been done recently on finding and classifying physical
invariants (see \eg [4]). However, so far the attention of researchers has
 rested almost
exclusively on
the special case where $\g_L=\g_R=\g$ and $k_L=k_R=k$ (perhaps the only
notable exception is level 1, the canonical example being the heterotic
string). Such an invariant will be called {\it symmetric}.
In the following section one such
approach [5], due to Warner and, independently, Roberts and Terao, shall be
described.

Perhaps the principal reason for the general absence of work on heterotic
 invariants is that the standard tools developed for the symmetric case
(\eg using simple currents [6], or automorphisms of the fusion rules of
extended algebras [7]) are not as easy to apply in the heterotic case. Those
algebraic techniques provide a very elegant  derivation of many symmetric
physical invariants, particularly those lying on infinite series, but
some of the so-called exceptional physical invariants are less tractible
from that perspective. Many exceptional invariants can be constructed
using conformal embeddings [8], but not all can. A major strength of the
Roberts-Terao-Warner lattice approach is that all invariants are placed
on an equal footing. Moreover, it is a complete method [9], and for small
ranks and levels is very practical. These qualities make it particularly suited
to the heterotic case, where we will find that there are only exceptional
physical invariants.

Their lattice method extends naturally to the heterotic case. However, the
resulting method will fail to find {\it any} nonzero
 modular invariants, unless the algebras
and levels involved satisfy excessively strong conditions. Fortunately, it is
possible to generalize their method, in two independent ways, so that
the completeness of the symmetric case is transferred into a completeness
for the heterotic case: this generalized lattice method will find all
possible heterotic partition functions (see Thm.E).

This paper is concerned with the search for heterotic physical invariants.
In Sec.2 we review the lattice approach of [5] for constructing
symmetric invariants. We generalize it in Sec.3.
In Sec.4 we analyse this new method. We find
a necessary condition on the algebras and levels required for the
existence of physical invariants. In Sec.5 we give explicit examples, and
use the analysis of Sec.4 to find all heterotic physical invariants of
smallest `total rank' (see eqs.(5.3)).
In the final section we provide a rigorous proof of the completeness of this
new method.

\bigskip \bigskip\centerline{{\bf 2. The lattice approach of
 Roberts-Terao-Warner}}\bigskip

In this section we will restrict our attention to the symmetric case
where $\g_L=\g_R=\g$ and $k_L=k_R=k$. We will review the method of
Roberts-Terao-Warner, changing their notation and presentation somewhat.
 But first we
will briefly introduce some of the notation involved with Kac-Moody
algebras (see [10,11] for details).

Let $g$ be a (simple) finite-dimensional Lie algebra of rank $n$, and let
$\alpha_1^{\vee},\ldots,\alpha_n^{\vee}$ be its simple coroots.
The span (over the integers \Z) of these coroots
is called the {\it coroot lattice} of $g$ and will be denoted by $M=M_g$.
These lattices $M$ are listed in the appendix for each choice of $g$. There
exist vectors $\beta_1,\ldots,\beta_n$ in the dual $M^*$
of the lattice $M$ satisfying $\beta_i\cdot \alpha_j^\vee=\delta_{ij}$;
these vectors span $M^*$ and are called the fundamental weights of $g$.
Let $\rho=\sum \beta_i$.

Each $g$ also has a set of colabels $a_i^\vee$, $i=1,\ldots,n$. The
number $1+\sum a_i^\vee$ is called the {\it dual Coxeter number} and is
denoted by $\h$. For each $k=0,1,2,\ldots$ define
$$P_{+}(g,k)\equi\bigl\{\sum_{i=1}^n m_i \beta_i\,|\, m_i\in \Z,\,\,
\,0\le m_i,\,\,\, \sum_{i=1}^n m_i a_i^\vee\le k\bigr\}\subset M^*.$$

Now consider the untwisted affine extension $\g=g^{(1)}$ of $g$. It turns out
that the properties of $\g$ can be expressed in terms of those of $g$.
The details will not be given here. Any integrable irreducible representation
of $\g$ is associated with a positive number $k$, called its level, and a
(horizontal) {\it highest weight vector} $\la\in P_{+}(g,k)$. For example, the
 singlet representation considered in (P3) corresponds to $\la=0$. A
 representation has the (normalized) character denoted by $\c_\la^{g,k}(u,z,\u)
$. Here, $u$ and $\u$ are complex numbers, Im$(\u)>0$,
and $z$ is a complex vector lying in
$${\bf C} \otimes M\equi \{\sum_i c_i\,x_i\,|\,c_i\in {\bf C},\sp x_i\in M\}.$$
(We will use that notation throughout this paper; ${\bf R}\otimes \l$ and
${\bf Q}\otimes \l$ are defined similarly, for any lattice $\l$.) The variable
$u$ is not
relevant for what follows and will be ignored (\ie set equal to 0).
Most writers consider the partition functions (1.1) to involve the
{\it restricted} characters $\c_\la^{g,k}(\u)$ where $z$ is also set equal
to 0. (However in [12] it is argued that for conformal field theories with
$c\ge 1$ --- which is the case of interest here --- these restricted
partition functions cannot carry enough information to specify the theory and
so $z$ should be retained.) In this paper we will retain the
vectors
$z$. Of course the restricted partition functions can be recovered at the
end by substituting in $z=0$.

We will only consider the case where the algebras $\g_L,\g_R$ are untwisted
affine algebras.
The approach of Roberts-Terao-Warner (henceforth RTW) starts from the Weyl-Kac
 character formula for these algebras. But before we can state it,
 we need the following definitions.

Given any positive definite lattice $\l$ and any vector $v\in \Q\otimes \l$,
 the translate $v+\l$ is called a {\it glue class}. The {\it theta series}
of that glue class is defined to be
$$\Theta\bigl(v+\l\bigr)(z|\u)\equi \sum_{x\in v+\l}\exp[\pi i \u\,x^2
+2\pi iz\cdot x].\eqno(2.1)$$
Here $\u\in {\bf C}$, and $z$ is a complex
vector lying in ${\bf C}\otimes \l$. Since $\l$ is positive definite, this
converges and in fact is analytic for all such $z$ and any $\u$ in the
upper half plane. Given any lattice $\l$ and positive number
$\ell$, we write $\l^{(\ell)}$ for the (positive definite) scaled lattice
 $\sqrt{\ell}\l$, and $\l^{(-\ell)}$ for the corresponding {\it negative
 definite} scaled lattice.

An {\it integral} lattice is one in which all dot products are integers.
An {\it even} lattice is both integral, and has only even norms.
An {\it odd} lattice is an integral lattice with at
least one odd normed vector. A {\it self-dual} lattice $\l$ is one which equals
 its dual $\l^*$. Finally, by a {\it gluing} $\l$ of $\l_0$, we mean that
$\l_0$
is a sublattice in $\l$ of finite index. Thus a gluing of $\l_0$ is
precisely any lattice which can be written as a finite disjoint union
of glue classes of $\l_0$.

For a discussion of glue classes and their uses, see for example [13].
The theta series of lattices are addressed for example in [14].

Now we return to the context of the algebra $g$. $M$ is its coroot lattice
and $W(g)$ is its Weyl group. Given a transformation $w\in W(g)$, we can
define its sign $\epsilon(w)=$det($w)\in \{\pm 1\}$.

The Weyl-Kac character formula can now be written as:
$$\eqalignno{\c_\lambda^{g,k}(z,\u)=&{\sum_{w\in W(g)} \epsilon(w) \,
\Theta\bigl({\lambda+\rho\over\sqrt{k+\h}}+M^{(k+\h)}\bigr)(\sqrt{k+\h}w(z)|\u)
\over D_g(z|\u)},&(2.2a)\cr D_g(z|\u)\equi &\sum_{w\in W(g)}\epsilon(w)\,
\Theta\bigl(
{\rho\over\sqrt{\h}}+M^{(\h)}\bigr)(\sqrt{\h}w(z)|\u).&(2.2b)\cr}$$

 Note that eq.(2.2$a$) allows us to define
$\c_\lambda^{g,k}$ for {\it any} $\lambda\in M^*$. However, it
can be shown
(see [10]) that for any $\lambda\in M^*$ and $k\geq 0$, either
$$\c_\la^{g,k}(z,\u)=0\eqno(2.3a)$$
for all $z$ and $\u$, or there exists a
unique $\lambda'\in P_{+}(g,k)$ and $\epsilon\in\{\pm 1\}$ such that
$$\c_\lambda^{g,k}(z,\u)=\epsilon \,\,\c_{\lambda'}^{g,k}(z,\u),\eqno(2.3b)$$
for all $z$ and $\u$.

The central idea of the RTW lattice method is as follows.

By $(M^{(\ell)};M^{(\ell)})$ we mean the $2n$-dimensional indefinite
lattice $M^{(\ell)}\oplus M^{(-\ell)}$; we will denote its vectors by
$(x_L;x_R)$ in the obvious way.
We are interested here in the choice $\ell=k+\h$. Let $\l$ be any
{\it even self-dual gluing} of $\l^{g,k}\equi (M^{(k+\h)};M^{(k+\h)})$. That
implies that
$$\l^{g,k}\subset \l =\l^*\subset \l^{g,k*}.$$ We will also write the vectors
of $\l$ in the form $(x_L;x_R)$.

Define the function $$\eqalignno{WZ_\l(g,k)(z_Lz_R|\u) &\equi
 \sum_{w_L,w_R \in W(g)} \epsilon(w_L) \epsilon(w_R)\sum_{(x_L;x_R)\in \l}
 \exp[\pi i \u x_L^2-\pi i \u^* x_R^2]\cdot&(2.4)\cr
&\cdot \exp[2\pi \sqrt{k+\h}
i(w_L(z_L)\cdot x_L-w_R(z_R)^*\cdot x_R)]/ D_g(z_L|\u)D_g(z_R|\u)^*&\cr}$$
Because $\l$ is a gluing of $\l^{g,k}$, eq.(2.2$a$) tells us we can write
$WZ_\l(g,k)(z_Lz_R|\u)$ as a sum of terms looking like
$$\c_{\la_L}^{g,k}(z_L,\u)\cdot\c_{\la_R}^{g,k}(z_R,\u)^*,\eqno(2.5)$$
for $\la_L,\la_R\in M^*$. Eqs.(2.3) now tell us that
$WZ_\l(g,k)(z_Lz_R|\u)$ can be written as a linear combination (over $\Z$) of
terms like that in eq.(2.5), with $\la_L,\la_R$ now lying in $P_{+}(g,k)$.
Also, because $\l$ is even, (1.2$a$) is satisfied, and because $\l$ is
self-dual, it can be shown (using Poisson's equation or more directly
 eq.(3.10$b$) in [14], together with Lemma 13.8 in [10]) that
(1.2$b$) holds.

The RTW method suggests we look at these $WZ_\l(g,k)$.
We know from the previous paragraph
that these functions are in the form of eq.(1.1), and that property (P1)
is satisfied. In general, (P2) and (P3) will not be. However, any linear
combination
$$\sum_i \ell_i \, WZ_{\l_i}(g,k)(z_Lz_R|\u) \eqno(2.6)$$
 which satisfies (P2) and (P3), will be a physical invariant
when each $\l_i$ is an even self-dual gluing of $\l^{g,k}$.

Roberts and Terao [5] expressed all known physical invariants corresponding
 to $g=A_1$ or $g=A_2$ in the form (2.6). This author has
shown in [9] that all (symmetric) physical invariants associated to any $g$,
 of any level $k$, must necessarily be expressible in this form.

\bigskip\bigskip \centerline{{\bf 3. Generalizing the RTW method}} \bigskip

The extension of eq.(1.1) to semi-simple algebras is trivial. By a {\it type}
$\T$ we mean the collection
$$\eqalignno{\T=&(\T_L;\T_R), \sp {\rm where}&(3.1a)\cr
\T_L=&\bigl(\{g_{L1},k_{L1}\},\{g_{L2},k_{L2}\},\ldots,\{g_{Ll},k_{Ll}\}\bigr),
&(3.1b)\cr\T_R=&\bigl(\{g_{R1},k_{R1}\},\ldots,\{g_{Rr},k_{Rr}\}\bigr).
&(3.1c)\cr}$$
Here each $g_{Li}$ and $g_{Rj}$ is a simple finite-dimensional Lie algebra,
and each $k_{Li}$ and $k_{Rj}$ is a nonnegative integer. For shorthand we
will write
$$\eqalignno{\l(\T_L)=&M_{g_{L1}}^{(k_{L1}+\h_{L1})}\oplus\cdots\oplus
M_{g_{Ll}}^{(k_{Ll}+\h_{Ll})},&(3.2a)\cr
P_{+}(\T_L)=&P_{+}(g_{L1},k_{L1})\times\cdots\times P_{+}(g_{Ll},k_{Ll}),
&(3.2b)\cr
\c_{\la_L}(\T_L)(z_L,\u)=&\c_{\la_{L1}}^{g_{L1},k_{L1}}(z_{L1},\u)
\cdots \c_{\la_{Ll}}^{g_{Ll},k_{Ll}}(z_{Ll},\u),&(3.2c)\cr}$$
with similar definitions for $\T_R$, where `$\oplus$' in (3.2$a$) denotes
the orthogonal direct sum of lattices and `$\times$' in (3.2$b$) denotes
the cartesian product of sets, and where $\la_L=(\la_{L1},\ldots,\la_{Ll})
\in \l(\T_L)^*$ and $z_L=(z_{L1},\ldots,z_{Ll})\in {\bf C}\otimes \l(\T_L)$.

Then (1.1) becomes
$$Z(\T)(z_Lz_R|\u)=\sum_{\la_L,\la_R}N_{\la_L\la_R}\,\c_{\la_L}(\T_L)(z_L,\u)
\,\c_{\la_R}(\T_R)(z_R,\u)^*,\eqno(3.3)$$
where the sum is over all $\la_L\in P_{+}(\T_L)$, $\la_R\in P_{+}(\T_R)$.

Because for any $g$,
$$P_{+}(g,0)=\{0\} \,\,\, {\rm and}\,\,\,\c_{0}^{g,0}(z,\u)=1, \eqno(3.4)$$
we can and usually will assume in (3.1) that no levels $k_{Li},k_{Rj}$
equal 0. When all $l+r$ levels are positive,
$\T$ is said to be a {\it positive type}. By definition, we will insist that
the type of any (nonconstant) invariant be positive.
We will call an invariant $Z$, or its type $\T$, {\it symmetric} if $l=r$ and,
 up to a
rearrangement of the indices, $(g_{Li},k_{Li}) =(g_{Ri},k_{Ri})$. If not
symmetric, they will be called {\it heterotic}.

Let $n_{Li}$, $n_{Rj}$ be the ranks of $g_{Li}$, $g_{Rj}$, and define
$n_L=n_{L1}+\cdots+n_{Ll}$, $n_R=n_{R1}+\cdots+n_{Rr}$.
By the {\it total rank} of an invariant (or its type) we mean the
number $n_L+n_R=n_{L1}+\cdots +n_{Rr}$.

In this section we are concerned with the search for physical invariants
of arbitrary type. The idea is to generalize the RTW approach, \ie
to find a class of lattices $\l$ and some function $f_\l(z_Lz_R|\u)$ with
the property that
$$\eqalignno{Wf_\l(z_Lz_R|\u)\equi&\sum_{w_{Li},w_{Rj}}\epsilon(w_{L1})
\cdots \epsilon(w_{Rr})\cdot&(3.5)\cr &\cdot
{f_\l(\sqrt{k_{L1}+\h_{L1}}w_{L1}(z_{L1}),\ldots,\sqrt{k_{Rr}+\h_{Rr}} w_{R1}
(z_{R1})|\u) \over D_{g_{L1}}(z_{L1}|\u)
\cdots D_{g_{Rr}}(z_{Rr}|\u)^*}&\cr}$$ (the sum is over all $w_{Li}\in
W(g_{Li})$, $w_{Rj}\in W(g_{Rj})$)
can firstly (i) be written as a linear combination of terms of the form
$$\c_{\la_L}(\T_L)(z_L,\u)\cdot \c_{\la_R}(\T_R)(z_R,\u)^*,\eqno(3.6)$$
 for $\la_L\in P_{+}(\T_L)$, $\la_R\in P_{+}(\T_R)$,
and secondly (ii) that for all $\l$ in that class of
lattices the function $Wf_\l$ has the correct modular behaviour (see
eqs.(1.2)).

The obvious extension of the RTW lattice method
to this more general situation is to let the lattices $\l$ in
(3.5) be even self-dual gluings of the indefinite lattice $$\l(\T)\equi
\bigl(\l(\T_L);\l(\T_R)\bigr),$$
and to define the function $WZ_\l(\T)(z_Lz_R|\u)$
as the obvious analogue of (2.4), by choosing $f_\l=\Theta(\l)$. However,
 by a well-known theorem
(see \eg [13]) such $\l$ could exist only when the ranks $n_{Li},n_{Rj}$
(which equal the dimensions of the coroot lattices $M_{g_{Li}},M_{g_{Rj}}$)
satisfy the congruence $n_L\equiv n_R$ (mod 8).
This excessively strong condition suggests this extension is not sufficiently
general.

There are two directions in which we will generalize the RTW method. We
will consider a more general $f_\l$, which will allow us to use {\it odd}
self-dual lattices as well as even ones. Secondly, we will use (3.4) to allow
us to use other types $\T'$ in our search for invariants of type $\T$.

Before we continue we should explicitly write down the modular behaviour
of the denominator $D_g$ of the Weyl-Kac formula, eq.(2.2).
$$\eqalignno{D_g(z|\u+1)=&\exp[\pi i \rho^2/\h]\, D_g(z|\u) &(3.7a)\cr
D_g(z/\u|-1/\u)=&\left( {\u \over i} \right)^{n/2} \exp[\pi i\h z^2/\u] \,
\exp[3\pi i \|\Delta_+\|/2] \,D_g(z|\u) &(3.7b)\cr}$$
where $\rho$ is defined in the previous section, and where
$\|\Delta_+\|$ denotes the number of positive roots of $g$. The numbers
$\rho^2$ and $\|\Delta_+\|$ for each $g$ are included in the appendix. The
proof
of (3.7$a$) follows immediately from the fact that the coroot lattice
$M$ is even for any Lie algebra $g$. The proof of (3.7$b$) is given
in Lemma 13.8 of [10].

Let $\l$ be any lattice (not necessarily self-dual), and let $u=(u_L;u_R),
v=(v_L;v_R)\in \R\otimes \l$ (${\bf R}\otimes \l$ is defined in the fourth
paragraph of Sec.2). A fairly general lattice function which is
reasonably well-behaved under $\u\rightarrow -1/\u$ is the following
function (similar to one found in [15] in the context of string theory, and
related to the theta functions of rational characteristic in [16]):
$$\eqalignno{A^{u,v}(\l)(z_Lz_R|\u)=&\sum_{(x_L;x_R)\in \l}
\exp[\pi i \u(x_L+u_L)^2-\pi i \u^*(x_R+u_R)^2]\cdot&(3.8a)\cr
&\cdot \exp[2\pi i\{(z_L+v_L)\cdot (x_L+u_L)-(z_R^*+v_R)\cdot
(x_R+u_R)\}].&\cr}$$
Then Poisson's eq. gives us (for integral $\l$)
$$\eqalignno{A^{u,v}(\l)(z_L/\u,z_R/\u|-1/\u)=&{\left( {\u \over i}
\right)^{n_L/2}
\left( {-\u^* \over i}\right)^{n_R/2} \over \sqrt{|\l|}} \exp[\pi iz_L^2/\u-
\pi iz_R^{*2}/\u^*] \cdot&\cr
&\cdot e^{2\pi i u\cdot v}\sum_{[g]\in \l^*/\l}A^{v+g,-u}(\l)(z_Lz_R|\u).
&(3.8b)\cr}$$
We would like to make the choice $f_\l=A^{u,v}(\l)$ in (3.5), which suggests
that we demand $\l^*=\l$, $v\equiv u$ (mod $\l$), and $-u\equiv v$ (mod
$\l^*$),
\ie that $\l$ be self-dual, $u=v$, and $2u\in \l$.

Eqs.(1.2$a$) and (3.7$a$) suggest that $A^{u,v}(\l)(z_Lz_R|\u+1)$ should
equal $A^{u,v}(\l)(z_Lz_R|\u)$ up to some constant factor, which in turn
suggests that we should require $(x+u)^2$ be independent (mod 2) of $x\in \l$.
This holds iff
$$x^2+2x\cdot u \equiv 0 \,\,\,({\rm mod\,\,\,}2)\,\,\, \forall x\in \l.
\eqno(3.9)$$
Given any integral $\l$, it is easy to show that there are always vectors
 $u\in \R\otimes \l$ satisfying (3.9): an example is
$$u_e={1\over 2}\sum_{i=1}^n e_i^2 e_i^* ,$$
where $\{e_i\}$ is any basis of $\l$ and $\{e_i^*\}$ is the corresponding
dual basis, which satisfies $e_i\cdot e_j^*=\delta_{ij}$. For self-dual $\l$,
the solution $u$ to (3.9) is unique, modulo $\l$. Moreover, for self-dual $\l$
any $u$ satisfying (3.9) must necessarily obey $2u\in \l$ and has the norm
$$u^2\equiv (n_L-n_R)/4 \,\,\,({\rm mod}\,\,\,2),\eqno(3.10)$$
where $(n_L,n_R)$ is the signature of $\l$ (eq.(3.10) is proved \eg in [17]).

By no means is it suggested that these choices of $\l,u,v$ are the most
general possible to yield modular invariant $A^{u,v}(\l)$ (that
would be false), but they turn out to be sufficient to span any other choice.

For any (not necessarily positive) type $\T$,
define $$\eqalignno{WZ_\l^u\bigl( \T \bigr)(z_Lz_R|\u)&\equi \exp[\pi i
 (n_R-n_L)/2]\sum_{w_{Li},w_{Rj}} \epsilon(w_{L1})\cdots\epsilon(w_{Rr})
\cdot&(3.11a)\cr &\cdot {A^{u,u}\bigl(\l\bigr)(\sqrt{k_{L1}+\h_{L1}}w_{L1}
(z_{L1}),\ldots,\sqrt{k_{Rr}+\h_{Rr}}w_{Rr}(z_{Rr})|\u)
\over D(\T_L)(z_L|\u)D(\T_R)(z_R|\u)^*}&
\cr D(\T_L)(z_L|\u)&\equi D_{g_{L1}}(z_{L1}|\u)
\cdots D_{g_{Ll}}(z_{Ll}|\u),&(3.11b)\cr}$$
with a similar formula for $D(\T_R)$,
where the Weyl automorphisms $w_{Li},w_{Rj}$ are as in (3.5), and
$n_L=\sum n_{Li}$, $n_R=\sum n_{Rj}$ (the factor in the front of (3.11$a$)
is introduced to avoid  awkward complex coefficients later).

The previous comments may be summarized in the following way:

\bigskip \noindent{\bf Theorem A}:\quad Let $\l$ be any self-dual gluing
of $\l(\T)$. Let $u\in \R\otimes \l$ be any vector satisfying eq.(3.9).
Then the function $WZ_\l^u\bigl(\T\bigr)(z_Lz_R|\u)$ has the following
properties:

\item{(i)} it can be written as a linear combination over $\Z$ of terms
of the form (3.6);

\item{(ii)} let $u'\in \R\otimes \l$ be any other vector satisfying eq.(3.9).
Then $$WZ^{u'}_\l(\T)=(-1)^{(u-u')^2}WZ^u_\l(\T);$$

\item{(iii)} $WZ^u_\l(\T)$ satisfies eq.(1.2$a$) iff
$$\sum_{i=1}^l {\rho^2_{Li} \over \h_{Li}} -\sum_{j=1}^r {\rho^2_{Rj} \over
\h_{Rj}} \equiv (n_L-n_R)/4 \,\,\,({\rm mod}\,\,\,2); \eqno(3.12a)$$

\item{(iv)} it satisfies eq.(1.2$b$) iff
$$\sum_{i=1}^l\|\Delta_{+Li}\|-\sum_{j=1}^r \|\Delta_{+Rj}\| \equiv n_L-n_R
\,\,\,({\rm mod}\,\,\,4).\eqno(3.12b)$$

\bigskip
This theorem defines a method for finding physical invariants of heterotic
type $\T$. At least for small dimensions $n_L$ and $n_R$ and levels $k_{Li},
k_{Rj}$, it is a straightforward task to find all self-dual gluings $\l_i$
of $\l(\T)$. Then as with eq.(2.6), we search through the linear combinations
of the resulting $WZ_{\l_i}^{u_i}(\T)$, to see when properties (P2) and (P3)
 are satisfied.

Note that we no longer require the self-dual lattice $\l$ to be even. Thus
we are freed from the severe restriction that $n_L\equiv n_R$ (mod 8). However
we are now limited by eqs.(3.12).

When $\l$ is even,  $u=0$ may be chosen. Since
$WZ_\l^0(\{g,k\};\{g,k\})=WZ_\l(g,k)$, we see then that  this new method
includes the RTW method as a special case. In fact, see Thm.D(i).

There is a second direction in which we can generalize the RTW method.
Consider any {\it positive} type $\T^+$.
Now consider any type $\T'$ of the form
$$\T'=\bigl(\{g_{L1}',0\},\ldots,\{g_{Ll'}',0\};\{g_{R1}',0\},\ldots,
\{g_{Rr'}',0\}\bigr),\eqno(3.13a)$$
where either $l'$ or $r'$ may be 0. Such a type, where all levels are 0, will
be called a {\it null type}. Let $\T=\T^++\T'$ denote the obvious combination
of these types:
$$\T^++\T'=\bigl(\{g_{L1},k_{L1}\},\ldots,\{g_{Ll'}',0\};\{g_{R1},k_{R1}\},
\ldots,\{g_{Rr'}',0\}\bigr). \eqno(3.13b)$$
We say that $\T^++\T'$ is the {\it augment} of $\T^+$ by $\T'$. We are
interested
in choosing $\T'$ so that $\T^++\T'$ satisfies eqs.(3.12), regardless of
whether
or not $\T^+$ alone does. The idea is that because of (3.4), the function
$WZ_\l^u(\T^++\T')$ for any self-dual gluing $\l$ of $\l(\T^++\T')$ will be
expressible as the sum of the terms in (3.5), and hence will actually be an
invariant of type $\T^+$.

This will be discussed further in the following sections. This approach
to construct physical invariants of type $\T^+$ by taking linear combinations
of the functions $WZ_\l^u(\T^++\T')$, will be called the {\it
generalized RTW method}. Examples will be be given in Sec.5.

Before we leave this section, let us express this more precisely and introduce
some useful notation. Consider any (not necessarily positive) type $\T$. Let
${\cal L}(\T)$ be the set of all pairs $(\l,u)$, where $\l$ is a self-dual
gluing of $\l(\T)$, and $\l,u$ satisfies eq.(3.9). Let ${\cal L}_{ev}(\T)$
consist of all {\it even} self-dual gluings of $\l(\T)$. There are finitely
many lattices in both ${\cal L}(\T)$
and ${\cal L}_{ev}(\T)$; ${\cal L}_{ev}(\T)$ will be empty
unless $n_L\equiv n_R$ (mod 8). Finally, let $\Omega^L(\T)$ be the space
consisting of all linear combinations (over ${\bf C}$) of the functions
$WZ_\l^u(\T)$, $\forall (\l,u)\in {\cal L}(\T)$, and let $\Omega^L_{ev}(\T)$
be the space spanned by all functions $WZ_\l^0(\T)$, for $\l\in {\cal L}_{ev}
(\T)$. We shall call these, respectively, the {\it lattice commutant} and
{\it even lattice commutant} of type $\T$.

Now let $\T^+$ be positive.
By $\Omega_W(\T^+)$ we mean the {\it Weyl-folded commutant} of type $\T^+$, \ie
the space of all modular invariant functions $Z(z_Lz_R|\u)$ of type $\T^+$.
The generalized RTW method involves getting at the Weyl-folded commutant,
 and hence the physical invariants, of type $\T^+$ by computing the (even)
lattice commutant of augmented type $\T^++\T'$, for some null type $\T'$.
We will show in Sec.6 that the (even) lattice commutant of type $\T^++\T'$
either equals $\{0\}$, or equals $\Omega_W(\T^+)$; we will also show that for
any positive $\T^+$ there exist null types $\T',\T''$ such that $\Omega^L
(\T^++\T')=\Omega_W(\T^+)$ and $\Omega^L_{ev}(\T^++\T'')=\Omega_W(\T^+)$;
and we will characterize all $\T',\T''$ for which these equalities hold.

The RTW method, on the other hand, may be equated with the set
$\Omega^L_{ev}(\T^+)$. For symmetric $\T^+$ this will always equal the
Weyl-folded commutant
of type $\T^+$ (see [9]), but for heterotic $\T^+$ it usually will not.

\bigskip\bigskip \centerline{{\bf 4. Analysis of the method}} \bigskip

This section begins the analysis of the completeness of the generalized
RTW method which will be concluded in Sec.6. It also includes a condition (4.3)
the type must satisfy in order that physical invariants of that type exist.

For the convenience of the reader,
in the appendix is listed $\rho^2$, $\h$, $\| \Delta_+ \|$ and $M$ for
each simple Lie algebra $g$. These quantities will be recurring
throughout this section.

In the following theorem we make use of a relation called {\it similarity},
 which is
discussed in some detail in [18]. It can be defined in the following way.
First, call two lattices $\l,\l'$, with bases $\{\beta_1,\ldots,\beta_n\},\,
\{\beta_1',\ldots,\beta_{n'}'\}$, {\it rationally equivalent} if their
associated quadratic forms $Q(x_1,\ldots,x_n)=(\sum_i x_i\beta_i)^2$ and
$Q'(y_1,\ldots,y_{n'})=(\sum_j y_j \beta_j')^2$ are transformable into each
other using linear maps ${\vec y}=A {\vec x}$ and ${\vec x}=B {\vec y}$,
where all $A_{ij}, B_{ij}\in {\bf Q}$.
Let $\l_1$ and $\l_2$ be two positive definite lattices of dimensions $n_1$
and $n_2$.
 Then we write $\l_1\sim \l_2$ iff $\l_1\oplus I_{n_2}$ and $\l_2\oplus
I_{n_1}$ are rationally equivalent. (Here and throughout this paper,
$I_n$ is the $n$-dimensional orthonormal lattice.) It turns out (see [18]
for details) that for direct sums of the root lattices and their scalings,
which is the case of interest here, similarities can easily be determined.
The similarities of the coroot lattices are also included in the appendix
below.

\bigskip \noindent{\bf Theorem B}: \quad The lattice method given in Thm.A
(\ie taking linear combinations of $WZ_\l^u(\T)$ for generating physical
 invariants requires that the following conditions on the (not
necessarily positive) type $\T=(\T_L;\T_R)$ must be satisfied:
$$\eqalignno{M(\T_L)\sim& M(\T_R); &(4.1a) \cr
\sum_{i=1}^l {\rho_{Li}^2 \over \h_{Li}}-\sum_{j=1}^r {\rho_{Rj}^2 \over
\h_{Rj}}&\equiv (n_L-n_R)/4\,\,\,({\rm mod}\,\,\,2);&(4.1b)\cr
\sum_{i=1}^l {\rho_{Li}^2 \over \h_{Li}+k_{Li}}-\sum_{j=1}^r {\rho_{Rj}^2
\over \h_{Rj}+k_{Rj}} &\equiv (n_L-n_R)/4\,\,\,({\rm mod}\,\,\,2).&(4.1c)
\cr}$$ \bigskip

Thus eqs.(4.1) are necessary for the lattice commutant $\Omega^L(\T)$ of type
$\T$ to contain physical invariants.
A (not necessarily positive) type $\T$ which satisfies eqs.(4.1) will be
called an {\it accessible type}; $\T$ will be called {\it weakly accessible}
if it satisfies eqs.(4.1$a,b$) but not necessarily (4.1$c$).

The condition in (4.1$a$) is precisely the statement [18] that the base lattice
$\l(\T)$ has self-dual gluings, in other words that the set ${\cal L}(\T)$
be nonempty. It implies among other things that the
determinant $|\l(\T)|$ must be a perfect square.

Eq.(4.1$b$) is just (3.12$a$), and is required for $WZ^u_\l(\T)$ to be
invariant under $\tau \rightarrow \tau+1$. By the ``strange'' formula of
Freudenthal-de Vries (or by explicit calculation using the values given
in the appendix) it can be seen that (3.12$b$) is satisfied whenever
(3.12$a$) is. Thus $WZ^u_\l(\T)$ will be an invariant iff (4.1$b$)
holds.

Eq.(4.1$c$) is the statement that the vector
$$x_u=-u+({\rho_{L1}\over \sqrt{k_{L1}+\h_{L1}}},\ldots;\ldots,{\rho_{Rr}
\over\sqrt{k_{Rr}+\h_{Rr}}})\eqno(4.2a)$$
must satisfy (3.9). If (4.1$c$) were not satisfied, then the vector $x_u$ could
lie in no $\l$, where $(\l,u)\in {\cal L}(\T)$. That would mean that no
$WZ^u_\l(\T)$, when
expanded into terms of the form of (3.3), would contain
$$\c_0(\T_L)(z_L,\u)\cdot\c_0(\T_R)(z_R,\u)^*,\eqno(4.2b)$$
and so no linear combination could possibly satisfy property (P3).

To help us
determine how overly restrictive eqs.(4.1) are, and hence how general the
method described in Sec.3 actually is, we will now give a condition
that the type of any physical invariant must satisfy.

\bigskip \noindent{\bf Theorem C}:\quad Let $\T^+$ be the type of some
{\it physical} invariant. Then the following condition must be satisfied:
$$\sum_{i=1}^l {\rho_{Li}^2 \over \h_{Li}+k_{Li}}-\sum_{j=1}^r {\rho_{Rj}^2
\over \h_{Rj}+k_{Rj}} \equiv \sum_{i=1}^l {\rho_{Li}^2 \over \h_{Li}}
-\sum_{j=1}^r {\rho_{Rj}^2 \over \h_{Rj}} \,\,\,({\rm mod}\,\,\,2).
\eqno(4.3)$$ \bigskip

Eq.(4.3) follows from invariance under $\u\rightarrow \u+1$ (\ie (1.2$a$)),
and the requirement (see property (P3)) that the invariant of type $\T$ must
contain the term in (4.2$b$) with nonzero coefficient. (It is proven in
[10] --- see also [14] --- that the products in (3.6) for $\la_L\in
P_{+}(\T_L)$ and $\la_R\in P_{+}(\T_R)$ are linearly
independent. They are linearly independent even when, in the symmetric case
considered by Roberts and Terao [5], we choose $z_L=z_R$ to be real.)

Again, it should be stressed that (4.3) must be satisfied by any type
which is realized by a physical invariant; it is {\it not} assumed that that
invariant is obtainable using the method introduced in Sec.3.

In the following section we will use Thm.C to help us find all heterotic
physical invariants of total rank $n_L+n_R<4$.

Note that a symmetric type automatically satisfies (4.3) (which is
to be expected, since any symmetric type has at least one physical invariant,
namely the diagonal one),
and also satisfies eqs.(4.1), \ie is also an accessible type.

For heterotic types, (4.3) is quite severe, but eqs.(4.1) are even more
so. A heterotic type $\T^+$ which has a physical invariant, need not be an
accessible type. When $\T^+$ is not accessible, its physical invariant(s)
cannot be obtained as a linear combination of $WZ^u_\l(\T^+)$. However, this
is not a serious problem, for the following result can be shown:

\bigskip \noindent{\bf Corollary C}: \quad Let $\T$ be any type.

\item{(i)} Then there exists a null type $\T'$ such that the augment
$\T+\T'$ is {\it weakly accessible}.

\item{(ii)} Suppose there is a physical invariant of type $\T$.
Then the augment $\T+\T'$ in (i) will in addition be {\it accessible}.
\bigskip

We will show in Sec.6 that the augment $\T+\T'$ in (ii) necessarily
satisfies $\Omega^L(\T+\T')=\Omega_W(\T)$. Incidently, there
are infinitely many null types $\T'$ that will work in (i) and (ii) (\eg add
arbitrary
numbers of copies of $\{A_1,0\}$ to either side of any $\T'$ which works).
 The proof of (i)
is not difficult, and (ii) is an immediate corollary of (i) and Thm.C.

The following result will simplify the completeness proof in Sec.6. It also
is useful in practical calculations, as will be illustrated in the next
section.

\bigskip \noindent{\bf Theorem D}:\quad Let $\T$ be any (not necessarily
positive) accessible type. Then

\item{(i)} if $n_L\equiv n_R$ (mod 8), then for {\it odd} $\l$ with $(\l,u)
\in {\cal L}(\T)$,
$WZ^u_\l(\T)$ can be written as the difference $WZ^0_{\l_1}(\T)-WZ^0_{\l_2}
(\T)$ for two {\it even} $\l_1,\l_2\in {\cal L}_{ev}(\T)$;

\item{(ii)} for $\T'=\bigl(\{A_1,0\};\,\,\bigr)$ or $\T'=\bigl(\,\,;
\{A_1,0\}\bigr)$, $\T+\T'$ is also accessible, and for any $(\l,u)\in
{\cal L}(T+T')$, $WZ^u_\l(\T+\T')$
equals $c\cdot WZ^{u'}_{\l'}(\T)$, for some $(\l',u')\in {\cal L}(\T)$, and
some constant $c$. \bigskip

To prove Thm.D(i), write $\l=\l_e\cup\bigl(\l_e+g\bigr)$, where $\l_e$ equals
 the set of all
even-normed vectors in $\l$ and $g^2$ is odd. Then define the gluings
$\l_1=\l_e\cup\bigl(\l_e+u\bigr)$ and $\l_2=\l_e\cup\bigl(\l_e+u+g\bigr)$.
The proof of (ii) is similar but a little
longer. First note $A_2^{(2)*}/A_2^{(2)}$ consists of 4 cosets: $[0]\equi 2\Z$;
$[1]\equi {1\over 2}+2\Z$; $[2]\equi 1+2\Z$; and $[3]\equi {3\over 2}+2\Z$,
using obvious notation. Write the cosets $x\in \l/\bigl(A_2^{(2)}\oplus \l(\T)
\bigr)$ as $[x_1,x_2]$, where $x_1=0,1,2,3$ and $x_2\in \l(\T)^*/\l(\T)$.
{\it case 1}: If $[2,0]\subset \l$ then $\l=\Z\oplus \l'$, where $\l'$ is a
self-dual gluing of $\l(\T)$. Write $u=(u_1,u')$, then $WZ^u_\l(\T+\T')=
2\cdot WZ^{u'}_{\l'}(\T)$. {\it case 2}: Otherwise, $\exists g\in \l(\T)^*
/\l(\T)$ such that $[1,g]\subset \l$. Without loss of generality (see
Thm.A(ii))
take $u=(0,u_2)$. Write $\l_2=\{x_2\in \l(\T)^*\,|\,(0,x_2)\in\l\}$. Then
$|\l_2|=4$ and $\l_2$ is integral. The order of $g$ in $\l_2$ is 4 (since
$[1]$ has order 4). {\it case 2a}: If $\l_2$ is an even lattice, then take
$\l'=\l_2\cup(\l_2+2g)$ and $u'=u_2+g$. {\it case 2b}: Otherwise there is
a vector $g'\in\l_2$ with odd norm. Then take $\l'=\l_2\cup(\l_2+2g)$ and
$u'=u_2+g'-g$.

Note that there are no conditions on $n_L,n_R$ in Thm.D(ii), so
the same conclusion holds for any $\T'=\bigl(\{A_1,0\},\ldots,\{A_1,0\};
\{A_1,0\},\ldots,\{A_1,0\}\bigr)$.
Thm.D(i) says that whenever $n_L\equiv n_R$ (mod 8),
$\Omega_{ev}^L(\T)=\Omega^L(\T)$. Of course, for $\T$ symmetric this forced
by the completeness proof in [9].

\bigskip\bigskip \centerline{{\bf 5. Examples: the physical invariants of
low rank}} \bigskip

In this section we will work out some explicit examples to help illustrate
 this method. In particular we will find all heterotic physical invariants
of total rank $n_L+n_R<4$.

Let us first ``solve'' (4.3) for the types of smallest total rank.
First consider positive types of total rank 1. These will look like
$\T^+=\bigl(\{A_1,k\};\,\,\bigr)$ or equivalently $\bigl(\,\,;\{A_1,k\}
\bigr)$, for some $k>0$. From the appendix we can write (4.3) as
$${{1\over 2}\over k+2}\equiv {{1\over 2}\over 2}\sp ({\rm mod}\sp 2),$$
\ie $k\equiv 0$ (mod 8($k+2)$). This has no solution for $k>0$. Thus
Thm.C tells us that there can be no physical invariant of total rank 1.

Similar but lengthier arguments apply to the other ranks. In particular
it can be shown that the only types of total rank 2 which satisfy (4.3) are
the symmetric types $\bigl(\{A_1,k\};\{A_1,k\}\bigr)$. There are exactly
6 solutions of total rank 3:
$$\eqalignno{\T_I=&\bigl(\{A_2,1\};\{A_1,4\}\bigr) &(5.1a)\cr
\T_{II}=&\bigl(\{A_1,1\},\{A_1,1\};\{A_1,4\}\bigr)&(5.1b)\cr
\T_{III}=&\bigl(\{C_2,1\};\{A_1,10\}\bigr) &(5.1c)\cr
\T_{IV}=&\bigl(\{G_2,1\};\{A_1,28\}\bigr)&(5.1d)\cr
\T_{V}=&\bigl(\{A_1,1\},\{A_1,2\};\{A_1,10\}\bigr)&(5.1e)\cr
\T_{VI}=&\bigl(\{A_1,3\},\{A_1,1\};\{A_1,28\}\bigr).&(5.1f)\cr}$$
All of these are inaccessible types --- \eg $\T_I$ fails to satisfy {\it any}
of eqs.(4.1). However the augments
$$\eqalignno{\T_I+\T_I'=&\bigl(\{A_2,1\},\{C_2,0\};
\{A_1,4\},\{A_3,0\}\bigr)&(5.2a)\cr
\T_{II}+\T_{II}'=&\bigl(\{A_1,1\},\{A_1,1\},\{G_2,0\},\{F_4,0\};
\{A_1,4\}\bigr)&(5.2b)\cr
\T_{III}+\T_{III}'=&\bigl(\{C_2,1\},\{G_2,0\};
\{A_1,10\},\{C_3,0\}\bigr)&(5.2c)\cr
\T_{IV}+\T_{IV}'=&\bigl(\{G_2,1\},\{C_2,0\};
\{A_1,28\},\{B_3,0\}\bigr)&(5.2d)\cr
\T_V+\T'_V=&\bigl(\{A_1,1\},\{A_1,2\},\{A_3,0\};
\{A_1,10\},\{D_5,0\}\bigr)&(5.2e)\cr
\T_{VI}+\T_{VI}'=&\bigl(\{A_1,3\},\{A_1,1\};\{A_1,28\},\{A_2,0\},\{D_7,0\}
\bigr)&(5.2f)\cr}$$
are all accessible.

 It can be shown (this will be discussed below)
that there is exactly one physical
invariant of types $\T_{I},$ $\T_{III}$, $\T_{IV}$, and $\T_{VI}$, and none of
types $\T_{II}$ and $\T_V$.
The rank 3 physical invariants are:
$$\eqalignno{Z_I=&\c_{00}\c_{0}^*+\c_{00}\c_4^*+\c_{10}\c_2^*
+\c_{01}\c_2^*&(5.3a)\cr
Z_{III}=&\c_{00}\c_0^*+\c_{00}\c_6^*+\c_{10}\c_3^*&\cr
&+\c_{10}\c_7^*+\c_{01}\c_4^*+\c_{01}\c_{10}^*&(5.3b)\cr
Z_{IV}=&\c_{00}\c_0^*+\c_{01}\c_6^*+\c_{00}\c_{10}^*
+\c_{01}\c_{12}^*&\cr
&+\c_{01}\c_{16}^*+\c_{00}\c_{18}^*+\c_{01}\c_{22}^*+
\c_{00}\c_{28}^*&(5.3c)\cr
Z_{VI}=&\{\c_0\c_0+\c_3\c_1\}\cdot\{\C_0+\C_{10}+\C_{18}+\C_{28}\}&\cr
&+\{\c_1\c_1+\c_2\c_0\}\cdot\{\C_6+\C_{12}+\C_{16}+\C_{22}\},&(5.3d)\cr}$$
where in the subscripts we used the Dynkin labels to represent the weights
$\la$, and the superscripts `$g,k$' have been dropped.

Thus the physical invariants of total rank $<4$ have been completely
classified; eqs.(5.3) contain the only heterotic ones, and [4] enumerated all
the symmetric ones, which are of type $\bigl(\{A_1,k\};\{A_1,k\}\bigr)$
(and total rank=2).

We will now use the generalized RTW method developed in Sec.3 to construct
the physical invariant $Z_I$ in (5.3$a$). The other physical invariants
in (5.3) can be constructed similarly.

Consider $\T_I$.
Because $\T_I$ does not satisfy $(4.1a)$, ${\cal L}(\T_I)=\{\}$ and $\Omega^L
(\T_I)=\{0\}$. That is one of the possible consequences of the type being
inaccessible, and is the reason that we must turn to the {\it accessible}
augment $\T_I''=\T_I+\T_I'$ given in (5.2$a$). We will construct the desired
self-dual lattice using {\it gluing}; for a review of the gluing theory
of lattices, see [13,18].

 The base lattice is $\l(\T_I'')=\bigl(A_2^{(4)}\oplus (A_1
\oplus A_1)^{(3)};A_1^{(6)}\oplus A_3^{(4)}\bigr)$, which can also be written
$(A_2^{(4)},I_1^{(6)},I_1^{(6)};I_1^{(12)},A_3^{(4)})$ using obvious
notation. This has the determinant $|\l(\T_I'')|=4^8\cdot 3^4$, so in order
for $\l$ to be a self-dual gluing, the glue group $\l/\l(T_I'')$ must have
order $\sqrt{4^8\cdot 3^4}=2^8\cdot 3^2=2304$.

A convenient basis for the dual $(A_2^{(4)})^*$ consists of the scaled
fundamental
weight $\beta_1/2$ and the scaled simple root $\alpha_1/2$. Let $\la_{m,n}$
denote the vector $m\beta_1/2+n\alpha_1/2$. Then $\la_{m,n}\in A_2^{(4)}$
iff $m\equiv 0$ (mod 12) and $n \equiv 0$ (mod 4). In this basis, $\rho/2=
\la_{3,-1}$.

Similarly, a convenient parametrization of $(A_3^{(4)})^*$ is $\mu_{p,q,r}=
p\beta_1/2+q\alpha_1/2+r\alpha_2/2$. $\mu_{p,q,r}\in A_3^{(4)}$ iff
$p\equiv 0$ (mod 16) and $q\equiv r\equiv 0$ (mod 4). The $\rho/2$ here
equals $\mu_{6,1,3}$.

Finally, $(I_1^{(6)},I_1^{(6)};I_1^{12)})^*$ can be parametrized by
the triples $(a,b,c)$ corresponding to the vector $(a/\sqrt{6},b/\sqrt{6};
c/\sqrt{12})$. That vector lies in $(I_1^{(6)},I_1^{(6)};I_1^{(12)})$
iff $a\equiv b\equiv 0$ (mod 6) and $c\equiv 0$ (mod 12). The scaled $\rho$
 for $(C_2^{(3)};A_1^{(6)})$ then is given by the triple $(2,1;1)$.

Note that by Thm.D(i) we may restrict our attention to even self-dual
gluings $\l$, since $n_L=n_R=4$. Because we want to find an invariant
satisfying (P3), we
may as well include the vector $x_0$ in $(4.2a)$ as a glue vector: hence our
first glue will be $g_1=(\la_{3,-1},2,1;1,\mu_{613})$. This glue has
order $24$. Of course it has even norm --- this is implied by (4.1$c$).
For our second glue we may choose any $g_2$ for which
$g_2\cdot g_1\in \Z$ and $g_2^2\in 2\Z$; an example is $g_2=(\la_{10},
1,1;0,\mu_{003})$, which has order 12. Similarly, $g_3=(0,3,3;6,0)$
and $g_4=(\la_{01},0,0;0,\mu_{010})$ have even norms and integral
dot products with all other glues. Their orders are 2 and 4, respectively.

Choosing $g_1,g_2,g_3,g_4$ as our glue generators, it is easy to see
that the group generated by them (mod $\l(\T_I'')$) has 2304 elements.
Thus the gluing $$\l=\l(\T_I'')[g_1,g_2,g_3,g_4]\,\equi\bigcup_{a,b,c,d\in \Z}
\bigl( ag_1+bg_2+cg_3+dg_4+\l(\T_I'')\bigr)$$
is even and self-dual, so lies in ${\cal L}_{ev}(\T_I'')$. It turns out to
have the partition
function $WZ_\l^0(\T_I'')=48Z_I$, and thus directly gives us one of the
physical invariants.

There are many different ways to show that eqs.(5.3) exhaust all the physical
 invariants of type $\T_I,\ldots,\T_{VI}$, including finding all relevant
lattice gluings (this is much easier than it sounds, thanks to the large Weyl
groups involved). For types of such small total rank and levels, the direct
 approach of explicitly calculating the $S$ and $T$ matrices corresponding
to the modular transformations $\u\rightarrow \u+1$ and $\u\rightarrow
-1/\u$ and finding all combinations of the characters invariant under them
also will work fine, but becomes totally unfeasible in more complicated
cases.

Incidently, refs.[5] both use the {\it shift} lattice construction
method. For the heterotic case (and perhaps also for the symmetric case),
the gluing method used above seems more efficient.

The four heterotic physical invariants obtained in (5.3) can also be
found using the technique of conformal embeddings [8]. For example, (5.3$a$)
comes from the level 1 diagonal invariant of $A_2$. However, the lattice
method found here is more general, and we can expect it to reveal more
solutions
than conformal embeddings for higher rank (see Thm.E below).

The {\it accessible} types of smallest total rank which satisfy (4.3) include
$$\eqalignno{\T_i=&\bigl(\{B_4,14 \};\,\,\bigr)&(5.4a)\cr
\T_{ii}=&\bigl(\{C_4,10 \};\,\,\bigr).&(5.4b)\cr}$$
Both of these are of total rank 4. It is not difficult to write down the other
heterotic rank 4 solutions to (4.3): there are (unfortunately) infinitely many
of type
$$\bigl(\{A_1,k\},\{A_1,k'\};\{A_1,k''\},\{A_1,k'''\}\bigr),$$
and approximately 50 others.

In the symmetric case, this lattice method is currently being used in a
computer program by Q. Ho-Kim [19] to find all symmetric physical invariants
of total rank 4, and levels up to around 35.

\bigskip\bigskip \centerline{{\bf 6. The proof of completeness}} \bigskip

In this section we prove the completeness of the {\it generalized RTW lattice
method}. Much of the notation used below is defined at the end of Sec.3.

In particular, consider any positive type $\T^+$. Let $\T'$ be any null type
for which the augment $\T=\T^++\T'$ is weakly accessible (see Cor.C(i)). That
means that $\T$ satisfies both eqs.(4.1$a,b$). The first of these says
that the lattice set ${\cal L}(\T)$ is nonempty, and the second says
that for any $(\l,u)\in {\cal L}(\T)$, the function $WZ_\l^u(\T)$
is modular invariant. It can be shown (see the proof of Thm.7.8 in [18]) that
${\cal L}_{ev}(\T)$
is nonempty iff both (4.1$a$) and $n_L+n_L'\equiv n_R+n_R'$ (mod 8) hold.
In that case we also have that for all $\l\in {\cal L}_{ev}(\T)$, $WZ_\l^0
(\T)$ is modular invariant.

\bigskip \noindent{\bf Theorem E}: \quad Consider any positive type $\T^+$. Let
$\T'$ be {\it any} null type for which the augment $\T^++\T'$ is weakly
accessible. Then the lattice commutant of type $\T^++\T'$ equals the
Weyl-folded
commutant of type $\T^+$:
$$\Omega^L(\T^++\T')=\Omega_W(\T^+).\eqno(6.1a)$$
Moreover, when the ranks satisfy $n_L+n_L'\equiv n_R+n_R'$ (mod 8), the
even lattice commutant equals the Weyl-folded commutant:
$$\Omega^L_{ev}(\T^++\T')=\Omega_W(\T^+).\eqno(6.1b)$$ \bigskip

Hence this says that the generalized RTW method succeeds in generating all
(heterotic) invariants, and in particular all physical ones. Although it
holds for any null type $\T'$ for which $\T^++\T'$ is weakly accessible,
clearly it is most advantageous to choose one of smallest rank, and if
possible one for which $n_L+n_L'\equiv n_R+n_R'$ (mod 8) also holds.
Of course, if $\T^+$ is already weakly accessible, then $\T'$ may be taken
to be empty: $\T'=\bigl(\,;\,\bigr)$.

In [9] we previously showed that the RTW lattice method spans the
commutant for symmetric $\T$. That argument was inspired by the analysis of
[20]. As much as possible, we will try in the following proof to stick
as close as possible to the machinery and notation developed in [9].

Note that Thm.A gives the containment $\Omega^L\subseteq \Omega_W$. To prove
Thm.E, we will augment the type $\T^++\T'$ by some copies of $\{A_1,0\}$.
This will allow us to consider the simpler case of even self-dual lattices;
by Thm.D, nothing is lost by using the resulting $\T_h$ instead of $\T^++\T'$.
Most of the proof is devoted to establishing a mapping, given in eqs.(6.7),
between the invariants of $\T_h$ and those of a certain symmetric type $\T_s$.
The
claim found below shows that each lattice partition function of $\T_s$ gets
sent by our mapping to a lattice partition function of $\T_h$.
We know [9] lattice partition functions span the commutant of $\T_s$, and
using our (linear, surjective) mapping we then get that the lattice partition
functions span that of $\T_h$.

\bigskip \noindent{\it Proof of Thm.E}\quad
Choose any number $n_L''\geq 0$ for which $n_L+n_L'+n_L''\equiv n_R+n_R'$
(mod 8), and let $\T''=\bigl(\{A_1,0\}^{n_L''};\,\,\bigr)$ be the null type
consisting of $n_L''$
copies of $\{A_1,0\}$. By Thm.D(ii), $\T_h\equi \T^++\T'+\T''$
will also be weakly accessible. Thanks to Thm.D, it will be more convenient
to work with $\T_h$ than $\T^++\T'$. We will return to $\T^++\T'$ in the final
paragraph of the proof.

Write $\T_h=(\T_L;\T_R)$. Our first goal will be to establish a connection
(see eqs.(6.7))
between invariants of the symmetric type $\T_s\equi(\T_L;T_L)+(\T_R;\T_R)$,
and invariants of the heterotic type $\T_h$. That will permit us to exploit the
result from [9] that lattice partition functions span the commutants of
symmetric types.

As was done in [9,20], we will consider instead the commutant built up from
the theta series of glue classes, rather than the one built up from
characters. The latter can be recovered from the former by
{\it Weyl-folding}, \ie summing over the Weyl group as in the numerator of
(2.2$a$). The details will be made clear below.

Define $\l_L=\l(\T_L)$ and $\l_R=\l(\T_R)$,
so that $\l(\T_h)=(\l_L;\l_R)$. For each $\la_L=(\la_{L1},\ldots,
\la_{Ln_L''}'')\in \l_L^*/\l_L$ and $z_L=(z_{L1},\ldots,z_{Ln_L''}'')
\in {\bf C}\otimes \l_L$, define the function
$$\eqalign{s_{\la_L}\bigl(\T_L\bigr)(z_L|\u &)\equi \left(
\prod_{i=1}^l {\Theta\bigl(
\la_{Li}+M_{g_{Li}}^{(k_{Li}+\h_{Li})}\bigr)(\sqrt{k_{Li}+\h_{Li}}z_{Li}|\u)
 \over D_{g_{Li}}(z_{Li}|\u)}\right) \cdot \cr
&\left( \prod_{j=1}^{l'} {\Theta
\bigl(\la_{Lj}'+M_{Lj}'{}^{(\h_{Lj}{}')}\bigr) (\sqrt{\h_{Lj}{}'}z_{Lj}'|\u)
\over D_{g_{Lj}'}(z_{Lj}'|\u)}\right) \cdot
\left(\prod_{\ell=1}^{n_L''} {\Theta\bigl(\la_{L\ell}''+A_1^{(2)}\bigr)
(\sqrt{2}z_{L\ell}''|\u)\over D_{A_1}(z_{L\ell}''|\u)}
\right) \cr}$$
and for $\la_R\in \l_R^*/\l_R$, $z_R\in {\bf C}\otimes \l_R$ define a function
 $t_{\la_R}\bigl(\T_R\bigr)(z_R|\u)$ similarly. All
terms $s_{\la_L}(\T_L)\,t_{\la_R}(\T_R)^*$ are linearly independent (this
follows from Thm.4.5 in [14]).

Define $\Omega_{th}(\T_h)$ to be the space of all {\it modular invariant}
linear combinations
$$Z=\sum N_{\la_L\la_R} \,s_{\la_L}(\T_L)\,t_{\la_R}(\T_R)^*, \eqno(6.2)$$
where the sum is over all $\la_L\in \l_L^*/\l_L,\la_R\in\l_R^*/\l_R$.
We will call these {\it theta-invariants} of type $\T_h$ to distinguish them
from the {\it character-}invariants considered elsewhere in this paper.
$N$ will be called the {\it coefficient matrix} of $Z$; its dimensions are
$|\l_L|\times |\l_R|$, where as usual we denote determinants of lattices
by $|\cdot |$.

There exist matrices $T_L$, $T_R$, $S_L$ and $S_R$ which describe how
$s_{\la_L}(\T_L)$ and $t_{\la_R}(\T_R)$, respectively, transform under $\u
\rightarrow\u+1$ and $\u\rightarrow-1/\u$. They are explicitly given in [9].
Here it suffices to remark that they are symmetric and unitary. The function
$Z$ in (6.2) is modular invariant iff both
$$\eqalignno{T_L^{\dag}\,N\,T_R &=N, \sp {\rm and} &(6.3a)\cr
S_L^{\dag}\,N\,S_R & =N. &(6.3b)\cr }$$

{}From the comments made at the beginning of this section, we know ${\cal
L}_{ev}
(\T_h)$ is nonempty, so there exists an even self-dual gluing $\l_0$ of
$(\l_L;\l_R)$ for which $WZ_{\l_0}^0(\T_h)$ is a modular invariant. In fact, an
identical calculation shows that the theta series of $\l_0$, divided by the
usual $D(\T_L)D(\T_R)^*$, also is modular invariant.
We will denote it by $Z_{\l_0}(\T_h)$: writing it out explicitly, we get
$$Z_{\l_0}\bigl(\T_h\bigr)(z_Lz_R|\u)=\sum_{(\la_L;\la_R)\in \l_0/(\l_L;\l_R)}
s_{\la_L}\bigl(\T_L\bigr)(z_L|\u)\,t_{\la_R}\bigl(\T_L\bigr)(z_R|\u)^*.
\eqno(6.4a)$$
It is in the form of (6.2) with coefficient matrix $N_{\l_0}$ given by
$$\left(N_{\l_0}\right)_{\la_L\la_R}=\left\{ \matrix{1&{\rm if \sp} (\la_L;
\la_R)\in \l_0\cr 0&{\rm otherwise}\cr}\right. . \eqno(6.4b)$$
Hence $Z_{\l_0}(\T_h)$ is a theta-invariant lying in $\Omega_{th}(\T_h)$.

Let $Z_L,Z_R$ with coefficient matrices $N_L,N_R$ be any theta-invariants
of type $(\T_L;\T_L)$ and $(\T_R;\T_R)$, respectively. Note that by (6.3),
$N_LN_{\l_0}N_R$ will be the coefficient matrix for a theta-invariant of type
$\T_h=(\T_L;\T_R)$.
This observation motivates the following discussion, designed to establish
the connection between $\T_h$ and $\T_s$ given in eqs.(6.7) below.

For any $\la_L,\la_L'\in \l_L^*/\l_L$, let $N^{\la_L\la_L'}_L(\T_L)$ denote
the coefficient matrix defined by:
$${A^{(\la_L;0),(\la_L';0)}(\l_L^D)\over D(\T_L)\,D(\T_L)^*}=
\sum_{\mu,\mu'\in \l_L^*/\l_L} \left(N^{\la_L\la_L'}_L(\T_L)\right)_{\mu\mu'}
s_\mu(\T_L)\,s_{\mu'}(\T_L)^*,\eqno(6.5)$$
where the function $A^{-,-}(\l)$ on the LHS is given by (3.8$a$) above, and
where $\l_L^D$ denotes the {\it diagonal gluing} of $(\l_L;\l_L)$
--- \ie the even, self-dual lattice
$$\l_L^D\equi \bigcup_{\la\in \l_L^*/\l_L}(\la;\la)+(\l_L;\l_L).$$
Define the analogous matrix $N^{\la_R\la_R'}_R(\T_R)$ similarly. The
dimensions of these two complex matrices are
$|\l_L|\times |\l_L|$ and $|\l_R|\times |\l_R|$, respectively.

Now choose any $k_L,k_L'\in (\l_L^{(2)})^*/\l_L^{(2)}$ and $k_R,k_R'\in
 (\l_R^{(2)} )^*/\l_R^{(2)}$. Define
the matrix $\{k_L,k_R,k'_L,k'_R\}_h$ by the matrix
product
$$\eqalignno{\{k_L,k_R,k_L',k_R'\}_h=&
\exp[-2\pi i(k_L\cdot k_L'+k_R\cdot k_R')]\cdot &\cr &\cdot
N^{\sqrt{2}k_L,\sqrt{2}k_L'}_L(\T_L)\,\,N_{\l_0}\,\,
N^{\sqrt{2}k_R,\sqrt{2}k_R'}_R(\T_R), &(6.6a)}$$
where by `$\sqrt{2}k_L$', etc. we mean the coset $\sqrt{2}k_L+\l_L\in \l_L^*
/\l_L$.
Because $N_{\l_0}$ is nonzero, an easy argument (see eq.(3.7$b$) in [9])
shows these $\{\cdots\}_h$
span the space of complex $|\l_L|\times |\l_R|$ matrices
(though they are not linearly independent). Moreover, from $(3.8b$) we
read off
$$\eqalignno{T_L^{\dag}\{k_L,k_R,k_L',k_R'\}_hT_R &=\{k_L,k_R,k_L'+k_L,k_R'+k_R
\}_h,&(6.6b)\cr
S_L^{\dag}\{k_L,k_R,k_L',k_R'\}_hS_R &=\{k_L',k_R',-k_L,-k_R\}_h. &(6.6c)\cr}$$
Note that eqs.(6.6$b,c$) are precisely the form obtained in the symmetric
types analysis of [9]. In particular, define
$$\{k_L,k_R,k_L',k_R'\}_s=\exp[-2\pi i(k_L\cdot k_L'+k_R\cdot k_R')]\,
N^{\sqrt{2}(k_L,k_R),\sqrt{2}(k_L',k_R')}_{LR}(\T_s),\eqno(6.6d)$$
where $N^{\sqrt{2}(k_L,k_R),\sqrt{2}(k_L',k_R')}_{LR}(\T_s)$ is defined
analogously to (6.5).
Then eqs.(6.6$b,c$) will still hold when $\{\cdots\}_h$ is replaced with
$\{\cdots\}_s$ (see eqs.(3.3) in [9]).

The point is the following. To any function of the form
$$Z_s=\sum_{k_L,k_R,k_L',k_R'}\alpha_{k_L,k_R,k_L',k_R'}\{k_L,k_R,k_L',k_R'
\}_s \eqno(6.7a)$$
we can assign the function
$$Z_h=\sum_{k_L,k_R,k_L',k_R'}\alpha_{k_L,k_R,k_L',k_R'}\{k_L,k_R,k_L',k_R'
\}_h. \eqno(6.7b)$$
(The well-definedness of this assignment will be discussed below.) If $Z_s$
is modular invariant, then so will be $Z_h$.

Moreover, {\it any} theta-invariant $Z_s\in \Omega_{th}(\T_s)$ and $Z_h\in
\Omega_{th}(\T_h)$ can be written in the forms of eqs.(6.7$a,b)$, respectively.
The proof of this is easy (see eq.(3.4$b$) in [9]), and is based on the
fact that a subgroup of finite index in the modular group fixes each
$\{\cdots\}_h$ and $\{\cdots\}_s$.

Thus eqs.(6.7) define a linear map from $\Omega_{th}(\T_s)$ into
$\Omega_{th}(\T_h)$. This map is well-defined (hence {\it onto}) because,
although $\{\cdots\}_s$
are not linearly independent, any choice of $\alpha$ for which the sum in
$(6.7a)$ is zero, will also give a zero sum in $(6.7b).$ The reason for this
is that the relations in (3.7$d$) of [9] hold in both the symmetric and
heterotic cases, and generate all linear dependencies in the symmetric case.
In general though there will be additional ones in the heterotic case, so
this map will usually not be {\it one-to-one}.

Let $\l_s\in {\cal L}_{ev}(\T_s)$, and define the lattice function $Z_{\l_s}
(\T_s)$ as in (6.4$a$). We proved in [9] that the theta-commutant of
type $\T_s$ will be spanned by these lattice functions. For any such $\l_s$,
let $Z_h(\l_s)$ denote the function in $(6.7b)$ assigned to $Z_{\l_s}(\T_s)$.
Then these $Z_h(\l_s)$ will necessarily span all of $\Omega_{th}(\T_h)$.

\bigskip \noindent{\bf Claim}:\quad For any $\l_s\in {\cal L}_{ev}(\T_s)$,
there exists a $\l_h\in {\cal L}_{ev}(\T_h)$ such that
$$Z_h(\l_s)=L\cdot Z_{\l_h}(\T_h),$$
for some integer $L>0$. \bigskip
\noindent{\it Proof of claim} \quad Let $N_h$ be the coefficient matrix
corresponding to $Z_h(\l_s)$. Then a simple calculation gives us, $\forall
\la_L\in \l_L^*/\l_L,$ $\la_R\in\l_R^*/\l_R$,
$$\left( N_h \right)_{\la_L\la_R}=\sum_{\la_L',\la_R'} \left( N_{\l_0}
\right)_{\la_L',\la_R'+2\la_R} \left( N_{\l_s} \right)_{\la_L,\la_R';\la_L',
-\la_R},\eqno(6.8)$$
where the sum is over all $\la_{L}'\in \l_{L}^*/\l_{L}$, $\la_R'\in \l_R^*
/\l_R$. But $N_{\l_0}$
and $N_{\l_s}$ look like (6.4$b$), so (6.8) can be simplified to the rule:

$(N_h)_{\la_L\la_R}$ equals the number of $\la_L',\la_R'$ for which both
$$\eqalignno{(\la_L';\la_R'+2\la_R)&\in \l_0, &(6.9a)\cr
(\la_L,\la_R';\la_L',-\la_R)&\in \l_s.&(6.9b)\cr}$$

Let $L=(N_h)_{00}$. Then $L\geq 1$ (since $\la_L'=\la_R'=0$ will always
work for $\la_L=\la_R=0$). By linearity of (6.9), we see immediately that
for each $\la_L,\la_R$, if $(N_h)_{\la_L\la_R}>0$, then $(N_h)_{\la_L\la_R}
=L$. Therefore $N_h/L$ has only the entries 0 and 1.

Define $\l_h\equi\{(\la_L;\la_R)\,|\, (N_h)_{\la_L\la_R}=L\}$. Then
$(\l_L;\l_R)
\subset \l_h$ (since $(N_h)_{00}=L$). We want to show that $\l_h$ is a
lattice. This will be so iff, for all $(\la_L;\la_R),(\la_L';\la_R')\in \l_h$,
 both
$$\eqalignno{(-\la_L;-\la_R)\in &\l_h,&(6.10a)\cr
(\la_L+\la_L';\la_R+\la_R')\in &\l_h &(6.10b)\cr}$$
hold. But this is immediate from the linearity of eqs.(6.9).

Thus $\l_h$ is a lattice, and $Z_h(\l_s)=L\cdot Z_{\l_h}(\T_h)$. Invariance
under
$\u\rightarrow \u+1$ immediately gives that $\l_h$ contains only even norms,
hence all its dot products are integers. Invariance under $\u\rightarrow
-1/\u$, and looking at the $(N_{\l_h})_{00}$ component, allows us to read
off that $|\l_h|=1.$

Therefore $\l_h$ is even and self-dual. \qquad QED to claim\bigskip

But any invariant in $\Omega_W(\T^+)$ can be written as one in $\Omega_{th}
(\T_h)$, simply by multiplying it by $1=\chi^0_{g_{L1}'}$ etc. and using
the Weyl-Kac formula to expand out the numerators of all characters into
theta functions. Weyl-folding the claim then tells us that
$\Omega_{ev}^L(\T_h)=\Omega^L(\T_h)=\Omega_W(\T^+)$. Thm.D(ii) now completes
the proof by saying $\Omega^L(\T^++\T')=\Omega^L(\T_h)$. \qquad QED to Thm.E

\bigskip\bigskip \centerline{{\bf 7. Comments}} \bigskip

In this paper we find a condition (4.3) which the algebras and levels must
satisfy in order for heterotic physical invariants to exist.
A lattice approach, called the {\it generalized RTW method}, for finding
heterotic physical invariants is proposed and analyzed. It will be summarized
 in the following paragraph. Using it, all heterotic physical invariants
of total rank $n_L+n_R<4$ are found --- see eqs.(5.3). We then prove in Thm.E
 that {\it any} heterotic physical invariant, of {\it any} type,
 can be obtained using this method.

The generalized RTW method:
suppose we are interested in finding physical invariants of (positive) type
$\T^+$. Then for any null type $\T'$ for which the augment $\T^+
+\T'$ is accessible,
find a self-dual gluing $\l$ of the base lattice $\l(\T^++\T')$. Then find any
$u$ satisfying (3.9) and compute the function $WZ^u_\l(\T^++\T')$, writing
it as a linear combination of terms looking like (3.6). Each of these
functions will be a modular invariant of type $\T^+$. Find the linear
 combinations of these
invariants which satisfy properties (P2) and (P3). Any linear combination
which does will be one of the desired physical invariants.

Examples of this method are
provided in Sec.5. For small ranks and levels, this method is extremely
practical [19].
There are other ways to find physical invariants (\eg conformal embeddings),
but a big advantage of this method is that it is complete: it will find all
of them.
But its greatest value may be theoretical, in that it offers
a convenient description of the entire commutant. Indeed, a logical first step
for classifying all physical invariants in a given class (see \eg [4,20])
involves understanding the commutant, and lattices could provide a valuable
tool for that. This is indicated in [21] by the transparency of a translation
 into the lattice language of the $A_1$ completeness proof, as well as in [9]
by the classification given there of the level 1 symmetric physical invariants.

It has been suggested in [21] that because the self-dual lattices
involved here may be {\it odd}, this generalized RTW approach
may be applicable to the study of coset theories. However, this question
has not yet been adequately investigated.

An intriguing use for heterotic invariants of small rank has been suggested
by C.S. Lam [22] as a means of reducing the rank of the effective gauge
group for heterotic strings. The idea is to factorize the partition function
of the theory: one factor will be a (small rank) physical invariant which
describes the real world, while the other factor will be a physical invariant
describing some `shadow world'. These factors must be chosen so that the
total central charge adds up to the correct numbers. For particle physics
we can ignore the shadow world and its invariant --- they would only be
relevant for gravitational considerations such as the cosmological constant
problem.

 \bigskip

This work is supported in part by the Natural Sciences and Engineering
Research Council of Canada. I would like to thank C.S. Lam for introducing
me to the problem of heterotic invariants, and Patrick Roberts for many
constructive criticisms. I have also benefited greatly
 from several
conversations with Quang Ho-Kim. Finally, the hospitality of the
Carleton mathematics department, where this paper was written, is much
appreciated.

\bigskip\bigskip \centerline{{\bf Appendix}} \bigskip

A number of relevant quantities for each simple Lie algebra $g$ is collected
below. By `$\sim$' we mean the similarity relation discussed in Sec.4, and
by $\{m_1,m_2,\ldots,m_k\}$ we mean the orthogonal lattice $I_1^{(m_1)}\oplus
\cdots \oplus I_1^{(m_k)}$. ($I_1$ is the 1-dimensional orthonormal lattice).

For $g=A_n$ ($n\geq 1$), $\rho^2=n(n+1)(n+2)/12$, $\h=n+1$, $\| \Delta_+ \|
=n(n+1)/2$, and $M=A_n$. $M\sim \{n+1,n+1,n+1\}$.

For $g=B_n$ ($n\geq 3$), $\rho^2=n(2n+1)(2n-1)/12$, $\h=2n-1$, $\| \Delta_+ \|
=n^2$, and $M=D_n$. $M\sim \{1\}$.

For $g=C_n$ ($n\geq 2)$, $\rho^2=n(2n+1)(n+1)/12$, $\h=n+1$, $\| \Delta_+ \|
=n^2$, and $M=A_1^n\equi A_1\oplus \cdots \oplus A_1$ ($n$ times). For $n$
even,
$M\sim \{1\}$ and for $n$ odd $M\sim \{2\}$.

For $g=D_n$ ($n\geq 4)$, $\rho^2=n(n-1)(2n-1)/6$, $\h=2(n-1)$, $\| \Delta_+
\|=n(n-1)$ and $M=D_n$. $M\sim \{1\}$.

For $g=E_6$, $\rho^2=78$, $\h=12$, $\| \Delta_+ \|=36$ and $M=E_6$.
$M\sim \{3\}$.

For $g=E_7$, $\rho^2=399/2$, $\h=18$, $\| \Delta_+ \|=63$ and $M=E_7$. $M\sim
\{2\}$.

For $g=E_8$, $\rho^2=620$, $\h=30$, $\| \Delta_+ \|=120$ and $M=E_8$. $M\sim
\{1\}$.

For $g=F_4$, $\rho^2=39$, $\h=9$, $\| \Delta_+ \|=24$ and $M=D_4$. $M
\sim \{1\}$.

For $g=G_2$, $\rho^2=14/3$, $\h=4$, $\| \Delta_+ \|=6$ and $M=A_2$. $M \sim
\{3,3,3\}$.

The root lattices $A_n$, $D_n$, $E_6$, $E_7$ and $E_8$
are described at some length in [13].

\bigskip\bigskip \centerline{{\bf References}} \bigskip

\item{[1]} A.A. Belavin, A.M. Polyakov and A.B. Zamolodchikov, {\it
 Nucl. Phys.} {\bf B241} (1984) 333

\item{[2]} D. Gepner and E. Witten, {\it Nucl. Phys.} {\bf B278} (1986) 493

\item{[3]} D.Gross, J.Harvey, E.Martinec and R.Rohm, {\it Nucl. Phys.} {\bf
B256} (1985) 253;

\item{} D.Gross, J.Harvey, E.Martinec and R.Rohm, {\it Nucl. Phys.}
{\bf B267} (1986) 75

\item{[4]} A. Cappelli, C. Itzykson and J.-B. Zuber, {\it Nucl. Phys.}
{\bf B280 [FS18]} (1987) 445;

\item{} A. Cappelli, C. Itzykson and J.-B. Zuber,
{\it Commun. Math. Phys.} {\bf 113} (1987) 1

\item{[5]} P. Roberts and H. Terao, {\it Int. J. Mod. Phys.} {\bf A7}
(1992) 2207;

\item{} N.P. Warner, {\it Commun. Math. Phys.} {\bf 130} (1990) 205

\item{[6]} A.N. Schellekens and S. Yankielowicz, {\it Int. J. Mod. Phys.}
 {\bf A5} (1990) 2903

\item{[7]} G. Moore and N. Seiberg, {\it Nucl. Phys.} {\bf B313} (1988) 16

\item{[8]} S. Bais and P. Bouwknegt, {\it Nucl. Phys.} {\bf B279} (1987) 561;

\item{}
A.N. Schellekens and N.P. Warner, {\it Phys. Rev.} {\bf D34} (1986) 3092

\item{[9]} T. Gannon, {\it WZW commutants, lattices, and level 1
partition functions} (Carleton preprint, 1992)

\item{[10]} V.G. Kac, {\it Infinite Dimensional Lie Algebras}, 3rd ed.
(Cambridge University Press, Cambridge, 1990)

\item{[11]} S.Kass, R.V. Moody, J. Patera and R. Slansky, {\it Affine Lie
Algebras, Weight Multiplicities, and Branching Rules} Vol.1 (University of
California Press, Berkeley, 1990)

\item{[12]} A.N. Schellekens, {\it Classification of ten-dimensional
heterotic strings} (CERN preprint TH.6325, 1991)

\item{[13]} J.H. Conway and N.J.A. Sloane, {\it Sphere packings,
Lattices and Groups} (Springer-Verlag, New York, 1988)

\item{[14]} T. Gannon and C.S. Lam, {\it J. Math. Phys.} {\bf 33} (1992) 871

\item{[15]} C.S. Lam, {\it Strings constructed from free Bose and Fermi
fields}, {\it Proceedings of Beijing String Workshop} (World Scientific, 1987)
(unpublished)

\item{[16]} D. Mumford, {\it Tata Lectures on Theta} Vol.1
(Birkhauser, Boston, 1984)

\item{[17]} J. Milnor and D. Husemoller, {\it Symmetric
Bilinear Forms} (Springer-Verlag, Berlin, 1973)

\item{[18]} T. Gannon and C.S. Lam, {\it Rev. Math. Phys}. {\bf 3} (1991) 331

\item{[19]} Q. Ho-Kim and T. Gannon, {\it The low level modular invariants
of rank 2 algebras} (work in progress)

\item{[20]} M. Bauer and C. Itzykson,
{\it Commun. Math. Phys.} {\bf 127} (1990) 617

\item{[21]} P. Roberts, {\it Whatever Goes Around Comes Around:
Modular Invariance in String Theory and Conformal Field Theory}, Ph.D. Thesis
(Institute of Theoretical Physics, Goteborg, 1992)

\item{[22]} C.S. Lam (private communication)

\end